\documentclass[italian,english]{article}
\usepackage[T1]{fontenc}
\usepackage[latin1]{inputenc}
\usepackage{color}
\usepackage{graphicx}
\usepackage{amssymb}

\makeatletter

 \newcommand{\lyxaddress}[1]{
   \par {\raggedright #1 
   \vspace{1.4em}
   \noindent\par}
 }

\usepackage{babel}
\makeatother
\begin{document}

\title{\textbf{Using long-baseline interferometric gravitational waves detectors
for high precision measures of the gravitational acceleration }}

\author{\textbf{Christian Corda}}

\maketitle

\lyxaddress{\begin{center}INFN - Sezione di Pisa and Università di Pisa, Via
F. Buonarroti 2, I - 56127 PISA, Italy\end{center}}

\lyxaddress{\begin{center}\textit{E-mail address:} \textcolor{blue}{christian.corda@ego-gw.it} \end{center}}

\begin{abstract}
A derivation of the optical axis lenght fluctations due by tilts of
the mirrors of the Fabry-Perot cavity of long-baseline interferometers
for the detection of gravitational waves in presence of the gravitational
field of the earth is discussed. By comparing with the typical tilt-induced
noises it is shown that this potential signal, which is considered
a weak source of noise, is negligible for the first generation of
gravitational waves interferometers, but, in principle, this effect
could be used for high precision measures of the gravitational acceleration
if advanced projects will achieve an high sensitivity. In that case
the precision of the misure could be higher than the gravimeter realized
by the Istituto di Metrologia {}``Gustavo Colonnetti''.
\end{abstract}

\lyxaddress{PACS numbers: 04.80.Nn, 04.80.-y, 04.25.Nx }

\section{Introduction}

The design and construction of a number of sensitive detectors for
gravitational waves (GWs) is underway today. We have some laser interferometers
like the VIRGO detector, being built in Cascina, near Pisa by a joint
Italian-French collaboration, the GEO 600 detector, being built in
Hanover, Germany by a joint Anglo-Germany collaboration, the two LIGO
detectors, being built in the United States (one in Hanford, Washington
and the other in Livingston, Louisiana) by a joint Caltech-Mit collaboration,
and the TAMA 300 detector, being built near Tokyo, Japan. We have
many bar detectors currently in operation too, and several interferometers
and bars that are in a phase of planning and proposal stages (for
the current status of gravitational waves experiments see \cite{key-1,key-2}).

The results of these detectors will have a fundamental impact on astrophysics
and gravitation physics. There will be lots of experimental data to
be analyzed, and theorists will be forced to interact with lots of
experiments and data analysts to extract the physics from the data
stream.

Detectors for GWs will also be important to verify that gravitational
waves only change distances perpendicular to their direction of propagation
and to confirm or ruling out the physical consistency of General Relativity
or of any other theory of gravitation \cite{key-3,key-4,key-5,key-6}.
This is because in the context of Extended Theories of Gravity we
know that some differences from General Relativity and the others
theories can be seen starting by the linearized theory of gravity
\cite{key-3,key-4,key-6}. 

In this paper we discuss on the possibility of an other use of long-baseline
interferometric GWs detectors, giving a derivation of the optical
axis lenght fluctations due by tilts of the two mirros of the Fabry-Perot
optical cavity in presence of the gravitational field of the earth.
By comparing with the typical tilt-induced noises it is shown that
this potential signal, which is considered a weak source of noise
\cite{key-7,key-8,key-9}, is negligible for the first generation
of GWs interferometers, but, in principle, this effect could be used
for high precision measures of the gravitational acceleration if advanced
projects will achieve an high sensitivity. In that case the precision
of the misure could be higher than the one of the gravimeter of the
Istituto di Metrologia {}``Gustavo Colonnetti'' (IMGC) \cite{key-10}.

\section{Using laser interferometers like gravimeters}

The coupling of angular tilts of the cavity mirrors with interferometer's
imperfections (like non perfect centering of the optical axis on the
two mirror rotation centers, interferometers asymmetries and input
laser geometry fluctations) has been long studied like a particular
noise source \cite{key-11,key-12,key-13}. In next Sections the effect
of the coupling of mirror tilts with gravitational earth field will
be analyzed, first like a source of noise for GWs interferometers,
comparing with other greater noises associated to mirror tilts (that
however do not limit the sensitivity of the interferometer) and showing
that in the first generation of interferometers this noise is negligible,
then the possibility of using this effect for high precision measures
of the gravitational acceleration with advanced projects will be discussed.

The possibility of using interferometers like high precision gravimeters
is well known in the scientific community. Actually falling corner
cube interferometers are being used for absolute measures of the gravitational
acceleration, being the best absolute instruments for this type of
misure, because they are based on first principle measures and require
no calibration \cite{key-10,key-14}. In particular the gravimeter
realized by the IMGC measures the gravitational acceleration with
a sensitivity of $10^{-8}ms^{-2}$ \cite{key-10}. The instrument,
based on symmetric motion, uses one mirror which is free falling in
a vacuum chamber. To detect the flight path, a laser interferometer
with a sensitivity of $10^{-9}m$ is used \cite{key-10}. In next
Sections of this paper it will be shown that advanced interferometers
for the detection of GWs could in principle achieve a better sensitivity
for measures of the gravitational acceleration.

\section{Effect of tilts in the frame of the local observer }

Because we are in a laboratory enviroment on earth, we typically use
the coordinate system in which the space-time is locally flat (see
refs. \cite{key-6,key-15,key-16,key-17}) and the distance between
any two points is given simply by the difference in their coordinates
in the sense of Newtonian physics. In this frame, called the frame
of the local observer, gravitational signals manifest themself by
exerting tidal forces on the masses (the mirror and the beam-splitter
in the case of a simple Michelson interferometer, see Figure 1, the
input and reflecting mirrors for a Fabry-Perot cavity, like the Virgo
interferometer, see Figure 2). %
\begin{figure}
\includegraphics{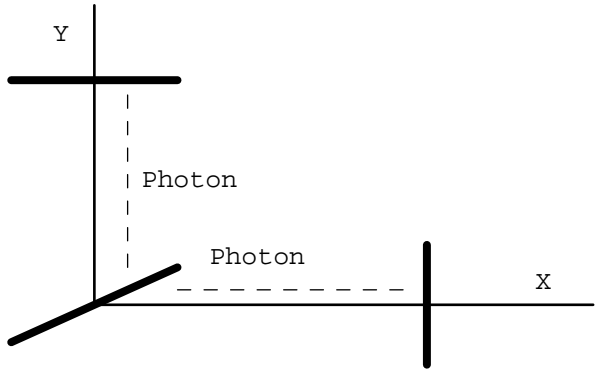}

\caption{a simple Michelson interferometer: photons can be launched from the
beam-splitter to be bounced back by the mirror}
\end{figure}
\begin{figure}
\includegraphics{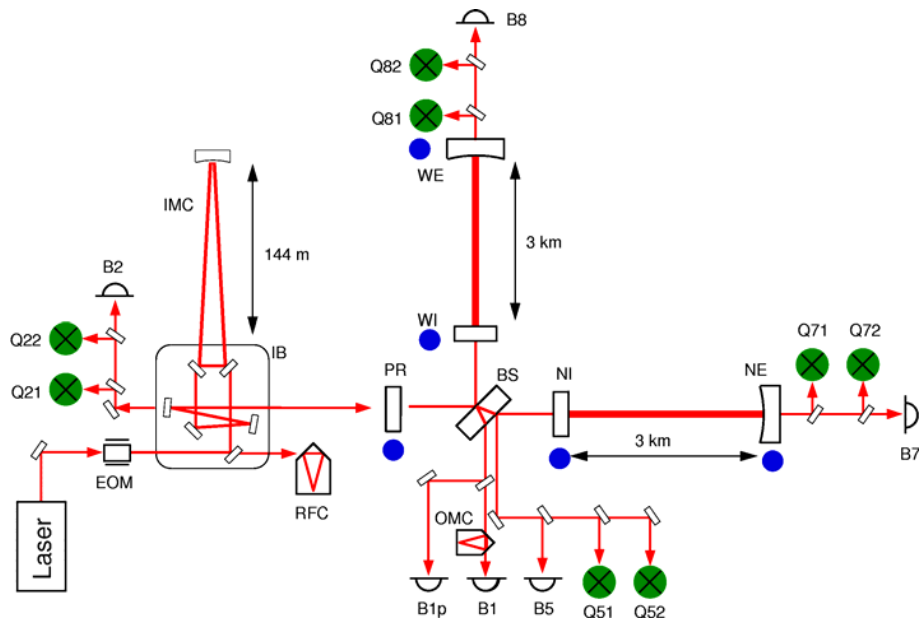}

\caption{Scheme of the Virgo interferometer}
\end{figure}

A detailed analysis of the frame of the local observer is given in
ref. \cite{key-15}, sect. 13.6. Here we remember only the more important
features of this frame:

the time coordinate $x_{0}$ is the proper time of the observer O;

spatial axes are centered in O;

in the special case of zero acceleration and zero rotation the spatial
coordinates $x_{j}$ are the proper distances along the axes and the
frame of the local observer reduces to a local Lorentz frame: in this
case the line element reads 

\begin{equation}
ds^{2}=(-dx^{0})^{2}+\delta_{ij}dx^{i}dx^{j}+O(|dx^{j}|^{2})dx^{\alpha}dx^{\beta};\label{eq: metrica local lorentz}\end{equation}

the effect of weak gravitational signals on test masses is described
by the equation for geodesic deviation in this frame

\begin{equation}
\ddot{x^{i}}=-\widetilde{R}_{0k0}^{i}x^{k},\label{eq: deviazione geodetiche}\end{equation}
where we have called $\widetilde{R}_{0k0}^{i}$ the linearized Riemann
tensor \cite{key-15}. 

Let us consider a Fabry - Perot cavity of lenght $L$ in the $x$
direction (i.e. we are assuming that the arms of our interferometer
are in the $x$ and $y$ axes), having the input mirror flat and the
far mirror convex of radius $R>L$ (like the Virgo interferometer,
see Figures 2, 3). 

\begin{figure}
\includegraphics{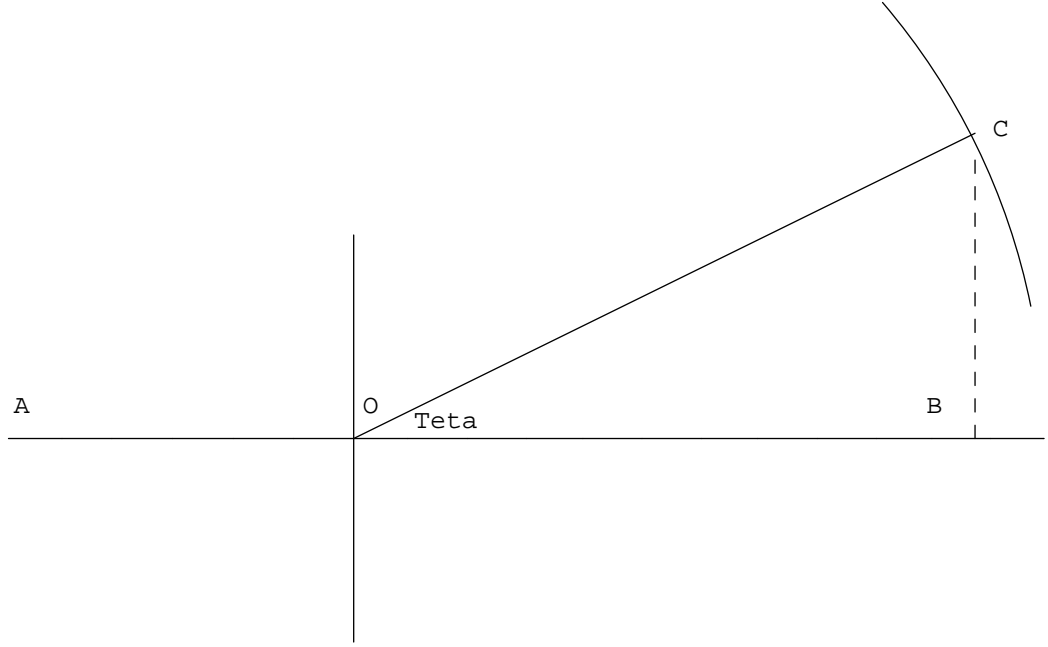}

\caption{the imput and far mirrors of a Fabry - Perot cavity, it is $AB=R$,
$OB=L$, $BC=d$ and $OC=L+a$}
\end{figure}
We are interested in vertical tilts generated by the earth gravitational
field (i.e. in the $z$ direction), thus, by putting the origin of
our coordinate system in the input mirror (Figure 3) we call $\theta$
the tilts of the second mirror around the horizontal axis from the
ideal position corresponding to the case of cavity axis lying on line
joining the mirror center of rotation. The geometrical effects on
the cavity are a first order displacement of the optical axis on the
mirror $2$ that we call $d$, and a second order variation of the
lenght $l$ of the cavity (ref. \cite{key-11} and Figure 3) given
by

\begin{equation}
d=(L+a)\theta\label{eq: d}\end{equation}

and \begin{equation}
l=L+d\theta=L+(L+a)\theta^{2}.\label{eq: l}\end{equation}

The more general interval in the frame of a local observer is \cite{key-6,key-15,key-16,key-17}\begin{equation}
ds^{2}=g_{00}dt^{2}+dx^{2}+dy^{2}+dz^{2},\label{eq: metrica osservatore locale generale}\end{equation}

(i.e. in this paper we work with $G=1$, $c=1$ and $\hbar=1$) where

\begin{equation}
g_{00}=-[1+2V(z)].\label{eq: g00}\end{equation}

Equation (\ref{eq: g00}) is the formula which gives the correspondence
between Newtonian theory and General Relativity and $V(z)$ is the
Newtonian potential of the earth \cite{key-6,key-15,key-16,key-17}.

Let us consider the interval for photons propagating along the $x$-axis
in absence of gravitational tilts \cite{key-6,key-15,key-16} \begin{equation}
ds^{2}=g_{00}dt^{2}+dx^{2}.\label{eq: metrica osservatore locale}\end{equation}
 We recall that the $y$ direction can be neglected because the absence
of the $y$ dependence in the metric (\ref{eq: metrica osservatore locale generale})
implies that photon momentum in this direction is conserved. We also
emphasize that photon momentum in the $z$ direction is not conserved,
for the $z$ dependence in eqs. (\ref{eq: g00}) and (\ref{eq: metrica osservatore locale generale})
(see ref. \cite{key-17} for details). Thus photons launched in the
$x$ axis will deflect out of this axis. But this effect can be neglected
because the photon deflection into the $z$ direction will be at most
of first order \cite{key-6,key-16}. Then, to first order, the $dz^{2}$
term can be neglected.

The condition for a null trajectory ($ds=0$) gives the coordinate
velocity of the photons \begin{equation}
v^{2}\equiv(\frac{dx}{dt})^{2}=1+2V(z=0),\label{eq: velocita' fotone in gauge locale}\end{equation}
which is a convenient quantity for calculations of the photon propagation
time between the the two the mirrors \cite{key-6,key-16}.

To first order the coordinate velocity is approximated by

\begin{equation}
v\approx\pm[1+V(z=0)],\label{eq: velocita fotone in gauge locale 2}\end{equation}

with $+$ and $-$ for the forward and return trip respectively. In
absence of gravitational tilts the coordinates of the input mirror
$x_{m1}=0$ and of the second mirror $x_{m2}=T=L$ do not change under
the influence of the gravitational field of the earth, thus we can
find the duration of the forward trip of the light as 

\begin{equation}
T_{01}=\int_{0}^{L}\frac{dx}{v}\label{eq:  tempo di propagazione andata gauge locale}\end{equation}

where $T=L$ is the transit time of the photon in absence of any gravitational
field, i.e. in a Minkowskian spacetime. To first order eq. (\ref{eq:  tempo di propagazione andata gauge locale})
can be rewritten as \begin{equation}
T_{01}\approx T-\int_{0}^{T}V(z=0)dx=T(1-V(z=0)).\label{eq:  tempo di propagazione andata1  in gauge locale}\end{equation}

In analogous way the propagation time in the return trip is

\begin{equation}
T_{02}=T-\int_{T}^{0}V(z=0)(-dx)=T(1-V(z=0))=T_{01}.\label{eq:  tempo di propagazione ritorno1  in gauge locale}\end{equation}
The sum of $T_{01}$ and $T_{02}$ give us the round-trip time for
photons traveling between the boundaries. Then we obtain the deviation
of this round-trip time (distance) from its unperturbed value $2T$
as

\begin{equation}
\delta(2T)=2T_{0}-2T=-2TV(z=0)).\label{eq: prima variazione}\end{equation}

Now let us suppose that, in presence of gravitational tilts, the second
mirror has a weak motion in the $z$ direction. In this case the mirror
$2$ rotates by an amount $\theta$ which corresponds to a parallel
displacement of the optical axis by the quantity

\begin{equation}
\delta(z)\equiv(T+a)\theta=d\label{eq: delta}\end{equation}

(see figure 3). Thus, to first order, we can write for the duration
of the forward trip

\begin{equation}
T_{*1}\approx T-(\int_{0}^{T}V(z)dx=T[1-(V(0)+\frac{\partial V(z)}{\partial z}\delta(z))].\label{eq:  tempo di propagazione andata2  in gauge locale}\end{equation}

In this way the time difference $\delta T_{1}=T_{*1}-T_{0}$ between
the tilted and unperturbed optical axes can be written as \begin{equation}
\delta T_{1}=Tg(T+a)\theta,\label{eq: dt1}\end{equation}
where 

\begin{equation}
g\equiv-\frac{\partial V(z)}{\partial z}\label{eq: acc  grav}\end{equation}

is the gravitational acceleration.

The duration of the return trip is computed as\begin{equation}
T_{*2}\approx T-(\int_{T}^{0}V(z))(-dx)=T[1-(V(0)+\frac{\partial V(z)}{\partial z}\delta(z)))]=T_{*1}.\label{eq:  tempo di propagazione ritorno2  in gauge locale}\end{equation}

Thus, the time difference $\delta T_{2}=T_{*2}-T_{0}$ between the
tilted and unperturbed optical axes for the return trip is the same
for the forward trip \begin{equation}
\delta T_{2}=Tg(T+a)\theta\label{eq: dt2}\end{equation}

and we have for the total variation of the round trip proper time
of a photon

\begin{equation}
\delta T=\delta T_{1}+\delta T_{2}=2Tg(T+a)\theta=2Tgd\label{eq: dt tot}\end{equation}

We have to recall that experimentalists compute the effect that this
variation of the round trip proper time generates in the variation
of the phase of the light of the laser in the cavity \cite{key-9,key-14},
thus

\begin{equation}
\delta\Phi=\Omega_{0}\delta T=2\Omega_{0}Tgd,\label{eq: fase}\end{equation}

is the signal which can be in principle computed with long-baseline
interferometric gravitational waves detectors \textbf{}and due to
gravitational tilts, where $\Omega_{0}=2\pi f_{0}$ and $f_{0}=10^{15}Hz$
in the Virgo interferometer.

For the other arm of the interferometer in which the mirrors are located
along the $y$ axis of our coordinate system the analysis is parallel
to the one above. 

In the case of the Virgo interferometer we have $L=T=3Km$ and, in
principle, we can compute distances (times) with a sensitivity of
$10^{-18}m$ and angles with a sensitivity of $10^{-9}rad$ \cite{key-7,key-8}.
In next section, by comparing with the typical tilt-induced noises,
it will be shown that precision measures of the gravitational acceleration
are in principle possible with long baseline interferometers with
a sensitivity of $10^{-9}ms^{-2}$ that is an order of magnitude higher
than the gravimeter realized by the IMGC.

\section{Typical error tilt noise in the Fabry-Perot cavity of laser interferometers}

The effect of coupling of mirror tilts with the gravitational earth
field is a first order effect, while, as it is well known \cite{key-7,key-8},
the lenght variation of a Fabry-Perot axis when mirrors rotate is
a second order effect. For an high precision alignment of the axis,
the gravitational coupling can overcome the other effects, thus it
could generate noise, or, from an other point of view, such an effect
could be put in evidence and measure the gravitational acceleration.
It is also known that, in principle, a small high-frequecies tilt
of angular amplitude $\theta\simeq10^{-9}rad$ can be detect \cite{key-7,key-8,key-9,key-13}.
But it has to be emphasized that, at least for the first generation
interferometers, the alignment of the cavity will not be perfect,
and the angular sensitivity will probabily grater than $10^{-9}rad$.
This is because the application of eq. (\ref{eq: l}) to gravitational
waves interferometers gives an intuitive simplification: interferometers
like Virgo potentially detect waves wich a frequency which falls in
the frequency-range $10Hz\leq f\leq10KHz$, \cite{key-6,key-7,key-8}
while typical spectrum of mirror rotations is domined by static or
slowly varyng components at frequencies well below $1Hz$. In the
range of interest for the GWs detection the Fourier transform of eq.
(\ref{eq: l}) is simply \cite{key-12}

\begin{equation}
\tilde{l}(f)\approx\tilde{d}\epsilon(f),\label{eq: l tilde}\end{equation}

where $\tilde{d}$ can be approximated with the root mean square of
long - term fluctations of the arm lenght $d$ defined by eq. (\ref{eq: d}).
Equation (\ref{eq: l tilde}) allows a direct comparision of the noise
due to coupling of small, high frequency mirror rotations with the
gravitational field or with the slowly varing mirror angle position.
Indeeed the estimed long term fluctation of the arm is of the order
of $\tilde{d}\approx10^{-5}$meters, which corresponds to an angular
amplitude $\theta\simeq10^{-5}rad$, while the gravitational coupling
generates an angular amplitude $\theta\simeq10^{-9}rad$. Thus the
coupling of mirror tilts with slowly varyng angle position is the
predominat effect. 

In order to performe a first measurement of the gravitational acceleration
with long baseline interferometers with a sensitivity higher than
the one of the gravimeter realized by the IMGC, the sensitivity should
be improved at a level better than $\theta\simeq10^{-9}rad$ which
corresponds to a a misure of the {}``gravitational coupling arm''
$\tilde{d}_{g}\approx10^{-9}$. This can be made in principe remembering
that long baseline interferometers recycle the power by using a particular
scheme that is shown in figure 2 (see also refs. \cite{key-7,key-8,key-18}).
The lenght of the equivalent cavity, which is constituited by the
recycling mirror (PR in figure two) and the Michelson interferometer,
is monitored and maintened on risonance by a suitable feedback system.
It is possible adding a small high-frequency sinusoidal modulaton
of tilts $\epsilon_{a}$ while a low-frequency misalignment $\tilde{d}_{c}$
is present, the lenght $\tilde{l}_{c}$ of the recycling cavity will
be affected by this geometrical coupling \begin{equation}
\tilde{l}_{c}=\tilde{d}_{c}\epsilon_{a},\label{eq: l tilde 2}\end{equation}

while the effect of the tilts due to the coupling with earth field
will be negligible because the short lenght of the cavity itself.
In this way, at the modulation frequency of $\epsilon_{a}$, the signal
will be useful for a good feedback to reduce the arm $\tilde{d}_{c}$.
Recolling that the sensivity to recycling cavity lenght fluctations
is extimed of the order $10^{-18}m/\sqrt{Hz}$ \cite{key-7,key-8,key-18}
 and the bandwidth of the feedback can be about $1Hz$, with a modulation
amplitude of $\epsilon_{a}\simeq10^{-8}rad$ a reduction up to $\tilde{l}_{c}\simeq10^{-10}m$
can be possible. Then, using the first generation parameters (even
if not in the first operation interferometers) in principle a measure
of $g$ with a precision of $10^{-9}-10^{-10}ms^{-2}$ is possible,
improving the one of the gravimeter realized by the IMGC. In advanced
projects of long-baseline interferometers for the detection of GWs
a better sensitivity of the order $10^{-19}m/\sqrt{Hz}$ is expected,
thus, using a modulation of $\epsilon_{a}\simeq10^{-7}rad$ (which
does not lower the interferometer performances) will be in principle
possible a measure of $g$ with a precision of $10^{-11}-10^{-12}ms^{-2}$.

\section{Conclusion remarks }

The possibility of an other use of laser interferometers for the detection
of gravitational waves with a derivation of the optical axis lenght
fluctations due by tilts of the mirrors of the Fabry-Perot cavity
in presence of the gravitational field of the earth has been discussed.
By comparing with the typical tilt-induced noises it has shown that
this potential signal, which is considered a weak source of noise,
is negligible for the first generation of gravitational waves interferometers,
but, in principle, this effect could be used for high precision measures
of the gravitational acceleration if advanced projects will achieve
an high sensitivity. In that case the precision of the misure could
be higher than the gravimeter realized by the Istituto di Metrologia
{}``Gustavo Colonnetti'' with a potential precision of $10^{-11}-10^{-12}ms^{-2}$.

\section*{Acknowledgements}

I would like to thank Salvatore Capozziello,  Maria Felicia De Laurentis
and Giancarlo Cella for helpful advices during my work. I have to
thank the referees for their interest in my work and for precious
advices and comments that allowed to improve this paper. The European
Gravitational Observatory (EGO) consortium has also to be thanked
for the using of computing facilities.


\begin{thebibliography}{10}
\bibitem{key-1}http://www.ligo.org/pdf\_public/camp.pdf
\bibitem{key-2}http://www.ligo.org/pdf\_public/hough02.pdf
\bibitem{key-3}Capozziello S - \textit{Newtonian Limit of Extended Theories of Gravity}
in \textit{Quantum Gravity Research Trends} Ed. A. Reimer, pp. 227-276
Nova Science Publishers Inc., NY (2005) - \foreignlanguage{italian}{also
in arXiv:}gr-qc/0412088 (2004)
\bibitem{key-4}Capozziello S and Troisi A - Phys. Rev. D \textbf{72} 044022 (2005)
\bibitem{key-5}Will C M \textit{Theory and Experiments in Gravitational Physics},
Cambridge Univ. Press Cambridge (1993)
\bibitem{key-6}Capozziello S and Corda C - Int. J. Mod. Phys. D \textbf{15} 1119
-1150 (2006)
\selectlanguage{italian}
\bibitem{key-7}Frasconi F and others (The Virgo Collaboration) - Class. Quant. Grav.
\textbf{21} 385-394 (2003)
\bibitem{key-8}Punturo M and others (The Virgo Collaboration) - 8th ICATPP proceedings
(2003)
\selectlanguage{english}
\bibitem{key-9}Saulson PR - {}`` Fundamentals of interferometric gravitational wave
detectors'' - World Scientific, Singapore - New Jersey - London -
Hong Kong (1994)
\selectlanguage{italian}
\bibitem{key-10}Germak A, Desogus S and Origlia C - Metrologia \textbf{39} \foreignlanguage{english}{(5)},
471-476 (2002)
\selectlanguage{english}
\bibitem{key-11}Rudiger A, Schilling R, Schnupp, L Winkler W, Billing H and Maischberger
K. - Opt. Acta 28 641 (1981) \foreignlanguage{italian}{}
\bibitem{key-12}Kawamura S and Zucker ME - Appl. Opt. \textbf{33}, 3912 (1994)
\bibitem{key-13}Calloni E, Di Fiore L, Di Sciascio G, Milano L, Rosa L, Stornaiolo
C - Phys. Lett. A \textbf{268} 235-240 (2000)
\bibitem{key-14}Private Communication with \foreignlanguage{italian}{the referees }
\selectlanguage{italian}
\bibitem{key-15}Misner CW, Thorne KS and Wheeler JA - {}``Gravitation'' - W.H.Feeman
and Company - 1973
\selectlanguage{english}
\bibitem{key-16}Rakhmanov M - Phys. Rev. D \textbf{71} 084003 (2005)
\bibitem{key-17}Landau L and Lifsits E - {}``Teoria dei campi'' - Editori riuniti
edition III (1999) \foreignlanguage{italian}{}
\bibitem{key-18}Acernese and others \foreignlanguage{italian}{(The Virgo Collaboration)}
- Astrpart. Physics \textbf{21} 1-22 (2004)
\end{thebibliography}
\end{document}